\documentclass[pra,aps,a4paper,twocolumn]{revtex4-2}

\usepackage{bm}
\usepackage{amsmath}
\usepackage{amsfonts}
\usepackage{amssymb}
\usepackage{color}
\usepackage{graphicx}
\usepackage[final]{pdfpages}
\usepackage{pgffor}
\usepackage{ulem}

\makeatletter
\AtBeginDocument{\let\LS@rot\@undefined}
\makeatother

\newcommand{\bs}[1]{\boldsymbol{#1}}
\newcommand{\mat}[4]{\left(\begin{array}{cc} {#1} & {#2} \\ {#3} & {#4} \end{array}\right)}

\newcommand{\ket}[1]{| #1 \rangle}
\newcommand{\bra}[1]{\langle #1 |}
\newcommand{\braket}[2]{\langle #1 | #2 \rangle}

\begin{document}

\title{Breaking of Coulomb blockade by macrospin-assisted tunneling}

\author{Tim \surname{Ludwig}$^{1}$, Rembert A. \surname{Duine}$^{1, 2}$}
\affiliation{$^1$Institute for Theoretical Physics, Utrecht University, Princetonplein 5, 3584 CC Utrecht, The Netherlands}
\affiliation{$^2$Department of Applied Physics, Eindhoven University of Technology, P.O. Box 513, 5600 MB Eindhoven, The Netherlands}

\begin{abstract}
A magnet with precessing magnetization pumps a spin current into adjacent leads. As a special case of this spin pumping, a precessing macrospin (magnetization) can assist electrons in tunneling. In small systems, however, the Coulomb blockade effect can block the transport of electrons. Here, we investigate the competition between macrospin-assisted tunneling and Coulomb blockade for the simplest system where both effects meet; namely, for a single tunnel junction between a normal metal and a metallic ferromagnet with precessing magnetization. By combining Fermi's golden rule  with magnetization dynamics and charging effects, we show that the macrospin-assisted tunneling can soften or even break the Coulomb blockade. The details of these effects---softening and breaking of Coulomb blockade---depend on the macrospin dynamics. This allows, for example, to measure the macrospin dynamics via a system's current-voltage characteristics. It also allows to control a spin current electrically. From a general perspective, our results provide a platform for the interplay between spintronics and electronics on the mesoscopic scale. We expect our work to provide a basis for the study of Coulomb blockade in more complicated spintronic systems.
\end{abstract}

\maketitle
\section{Introduction}
To compete with modern electronics, systems of spintronics---the spin analog of electronics---are becoming smaller. In turn, mesoscopic effects become more important. On the one hand, this complicates the description of spintronic effects. On the other hand, however, it opens up new ways to investigate and manipulate spintronic systems. Here, we demonstrate how the Coulomb blockade---a prominent effect of mesoscopic physics---can be used to measure magnetization dynamics via a system's current-voltage characteristics.

\begin{figure}[h]
\begin{center}
\includegraphics[width=0.4\textwidth]{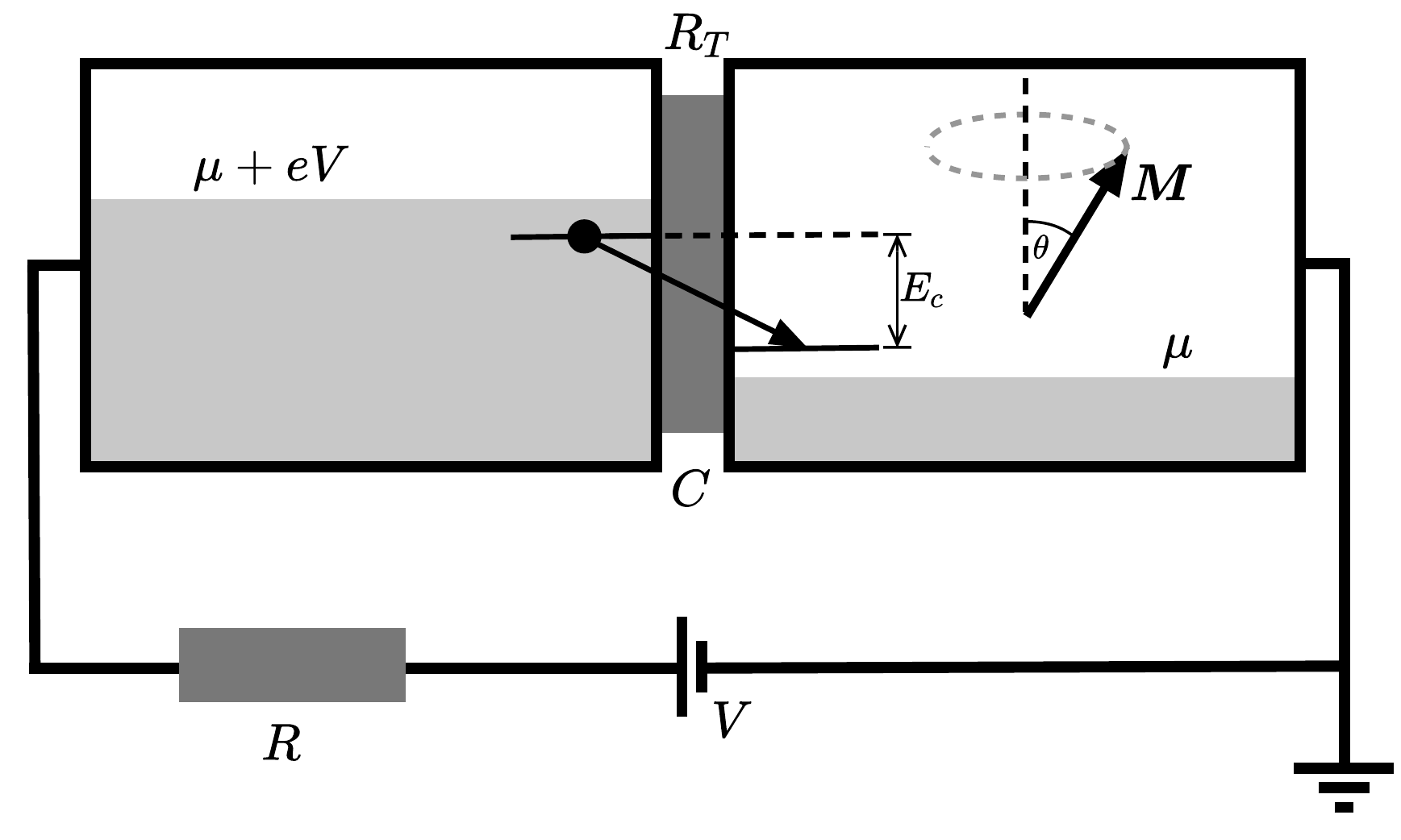}
\end{center}
\caption{We consider a single tunnel-junction between a normal metal (left lead) and a metallic ferromagnet (right lead) with a magnetization in a steady state precession. Separating two metallic systems, the tunnel-junction forms a capacitor (capacitance $C$).  For $R \gg R_K = h/e^2$, an electron, when tunneling, changes the charge on the capacitor by $e$ and, in turn, it looses the charging energy ($E_c = e^2/2C$) to its electrostatic environment \cite{ingold1992charge, nazarov2009quantum}. For low voltage ($eV<E_c$) and low temperature ($k_b T \ll E_c$), this energy loss usually forbids charge transport (Coulomb blockade). However, we show that the precessing magnetization can break the Coulomb blockade by assisting electrons in tunneling.} \label{fig: system}
\end{figure}

We consider a single tunnel-junction between a normal metal and a metallic ferromagnet; see Fig. \ref{fig: system}. A tunnel junction is a thin insulating layer which separates two metallic systems (leads) from each other. Classically, a tunnel junction forms a capacitor, as the insulating layer separates the two metallic systems by a small distance. Quantum mechanically---as the name "tunnel junction" suggests---electrons can tunnel through the insulating layer. The classical and quantum perspectives are related: when an electron tunnels through the junction, it changes the charge on the capacitor and, in turn, it changes the electrostatic Coulomb energy stored in the capacitor \cite{ingold1992charge, nazarov2009quantum}. If an electron does not have enough energy to compensate the cost in Coulomb energy, then the tunneling is blocked; this is the Coulomb blockade \cite{ingold1992charge, nazarov2009quantum}.

The energy to overcome Coulomb blockade can come from thermal fluctuations, from the voltage source, or---in the present case---from the magnetization dynamics. To focus on the role of the magnetization dynamics, we consider the limit of zero temperature. For simplicity, we assume the magnetization to precess in a steady state and we use the macrospin approximation; that is, we describe the magnetization as a single vector $\bs{M}$. A precessing macrospin (magnetization) acts as a time-dependent magnetic field for electrons and---as a special case of adiabatic pumping \cite{adiabaticpumping1, adiabaticpumping2}---pumps spin-polarized electrons into adjacent leads \cite{PhysRevLett.88.117601, brataas2002spin, RevModPhys.77.1375}. For a tunnel-junction, this means that a precessing macrospin can assist electrons in tunneling \cite{ludwig2019current}; see also \cite{zhang1997quenching, bender2019quantum}. 

In this article, we study the competition between Coulomb blockade and macrospin-assisted tunneling. This places our work into the emerging field of mesoscopic spintronics. Other topics in this field include, for example, the study of noise in spintronics \cite{aliev2018noise, cascales2012controlling, kamra2017spin, kamra2016super, PhysRevLett.114.176806, virtanen2017spin, ludwig2019current, PhysRevB.102.054440}, the mesoscopic Stoner instability \cite{kurland2000mesoscopic, sothmann2012mesoscopic, saha2012quantum,  burmistrov2020mesoscopic}, or spintronics with quantum dots \cite{konig2003interaction, braun2004theory, bender2010microwave, PhysRevB.87.155428, hell2015spin, gergs2018spin} which is closely related to our work. Here, we show that the macrospin-assisted tunneling can provide enough energy to overcome the Coulomb blockade. Thereby, the macrospin-assisted tunneling softens or even breaks the Coulomb blockade and, as a result, the system's current-voltage characteristics reveals the macrospin dynamics.

For slow precession, the macrospin-assisted tunneling softens the Coulomb blockade; see Fig. \ref{fig: softening}. For strong precession, the macrospin-assisted tunneling breaks the Coulomb blockade; see Fig. \ref{fig: breaking}. In both cases, the macrospin dynamics governs the flow of charge current. Read in reverse: the charge current can be used to measure the magnetization dynamics; see Figs. \ref{fig: softening} and \ref{fig: breaking}. To be more explicit, we choose the macrospin's precession axis as the $z-$direction. The precessing macrospin is then described by $\bs{M} =  M (\sin \theta \cos \phi, \sin \theta \sin \phi, \cos \theta)$, where polar angle $\theta$ and precession frequency $\dot \phi$ are constant. Roughly speaking: the polar angle $\theta$ can be inferred from the tunnel-junction's differential conductance; while the precession frequency $\dot \phi$ can be inferred from the combined information of conductance and differential conductance. Alternatively, one can infer $\theta$ and $\dot \phi$ from the position of kinks in the current-voltage characteristics; see Figs. \ref{fig: softening} and \ref{fig: breaking}.

We close the introduction by relating our work to Ref. \cite{PhysRevB.90.174431}, where the effect of an electromagnetic environment was studied for a magnetic tunnel junction with one precessing magnetization. Using $P(E)$-theory, they derive a general expression for the charge current, which should also cover our result \footnote{Unfortunately, there is a typographical error in their equation (8): what they call $V_s$ and $T_s(\theta)$ (where $s = \pm$) should always come with opposite signs; that is, $V_+$ should come with $T_-(\theta)$ and $V_-$ should come with $T_+(\theta)$. Once that is corrected, their equation (8) includes our equation (5) as a limiting case.}. However, their focus is on open circuits, where no charge current can flow and a voltage build-up is predicted instead. In open circuits---with always vanishing charge current---the Coulomb blockade and its breaking cannot be seen directly. In contrast to their work, we use a simple Fermi's golden rule approach, treat the electrical environment classically, and focus on closed circuits. While less general, our simple approach makes the physics particularly transparent. So, we can easily see how macrospin-assisted tunneling breaks the Coulomb blockade. In retrospect, we can even identify the results of Ref. \cite{PhysRevB.90.174431} as important indirect signs of the breaking of Coulomb blockade by macrospin-assisted tunneling.

\section{The Model}The breaking of Coulomb blockade by macrospin-assisted tunneling can be found within a simple model: the electrons are described quantum mechanically by single-particle Hamiltonians; whereas, the Coulomb energy is taken into account on the classical level.

The magnetic right lead is described by $H_r = \sum_{\rho \sigma \sigma'} \ket{\rho \sigma} h_{r, \rho}^{\sigma \sigma'} \bra{\rho \sigma'}\, $; where $\sigma$ and $\sigma'$ denote the spin in $z$-direction (the magnetization's precession axis), $\rho$ denotes the right-lead states with corresponding energy $\epsilon_\rho$, and $h_{r, \rho}  = \epsilon_\rho - \bs{M} \boldsymbol{\sigma}/2$ with the vector of Pauli-matrices $\bs{\sigma}$ and, for simpler notation, the magnetization length $M$ includes all constants \footnote{In detail, $M = \hbar \mu_0 \mu_B g \bar M$, where $\bar M$ is the real magnetization and we assumed the external magnetic field to be negligible, $\hbar$ is the reduced Planck constant, $\mu_0$ is the vacuum permeability, $\mu_B=e \hbar /2 m_e$ is the Bohr magneton, and $g$ is the Lande g-factor.}. 
The left lead is described by $H_l = \sum_{\lambda \sigma} \ket{\lambda \sigma} \epsilon_\lambda \bra{\lambda \sigma}\, $; where $\lambda$ denotes the left-lead states with the corresponding energy $\epsilon_\lambda$.
We assume spin-conserved tunneling, described by $H_t = \sum_{\lambda \rho \sigma}  \ket{\rho \sigma} t_0 \bra{\lambda \sigma} + h.c.$\ , where $t_0$ are the tunneling amplitudes between states $\rho$ and $\lambda$; for simplicity, we disregard the state-dependence of $t_0$. 

In addition to the Hamiltonian, we need to specify the distribution functions. We assume the tunneling events to be rare, such that local equilibrium is re-established before each tunneling event. Then, the electrons are distributed according to the Fermi distribution $f(\epsilon) = 1/[\exp[(\epsilon - \mu)/T]+1]$, where we assume the chemical potential $\mu$ and the temperature $T$ to be equal in both leads. We emphasize, however, that $\mu$ is only the \textit{chemical} potential---not the \textit{electrochemical} potential.

\section{The Coulomb blockade regime}
The (electrostatic) Coulomb energy is taken into account in addition to the single-particle contributions. In the circuit, Fig. \ref{fig: system}, there are two different scales for the electrostatic energy: first, there is $eV$, which is the work done by the voltage source when one electron is pumped from one side to the other; second, there is $Q^2/(2C)$, which is the energy stored in the capacitor with capacitance $C$ and charge $Q$. Which of these energy scales is relevant for the tunneling of electrons? As discussed in Ref. \cite{ingold1992charge}, this strongly depends on the tunnel junction's environment---in particular, it depends on the environmental resistance $R$: for a small resistance $R\ll R_K$, the (ideal) voltage source fixes the charge on the capacitor to $Q= CV$, such that the energy stored in the capacitor remains fixed and the tunneling is governed by the voltage source; for large resistance $R \gg R_K$, in contrast, the voltage source cannot immediately restore the charge on the capacitor, such that the capacitor's energy governs the tunneling. The natural scale separating these two cases is the resistance quantum (von Klitzing constant) $R_K = h/e^2$ \cite{ingold1992charge}. In the following, we focus on $R \gg R_K$ and the limit of zero temperature $T= 0$, which puts the system into the Coulomb blockade regime.

An electron can only tunnel if it has enough energy available. When one electron tunnels, the charge on the capacitor $Q$ is changed by $-e$ for left-to-right tunneling or by $+e$ for right-to-left tunneling. So, the change in electrostatic energy is $\Delta E_{el} = Q^2/2C - (Q\mp e)^2/2C$. Assuming tunneling events to be rare, the capacitor is recharged to $Q=CV$ before each tunneling event. In turn, we find $\Delta E_{el} = \pm eV - E_c$ with the charging energy $E_c = e^2 / 2C$. If the applied voltage is too small ($\Delta E_{el} <0$), we enter the regime of Coulomb blockade, where electrons cannot tunnel unless the missing electrostatic energy is supplied in a different way \cite{ingold1992charge, nazarov2009quantum}.
At $T=0$, the missing energy cannot come from thermal activation. However, the precessing macrospin can assist electrons in tunneling \cite{ludwig2019current} and, thereby, it provides the missing energy. 

\section{Macrospin-assisted tunneling}
As a first step, we determine the tunneling rate between states in the left lead and states in the right lead. We assume the tunnel-coupling to be a weak perturbation, such that we can use Fermi's golden rule.

Before Fermi's golden rule can be applied, we have to change to the magnetization's rotating frame of reference, such that the leads' Hamiltonians become time independent. So, following Ref. \cite{ludwig2019current}, we apply a transformation $U(t) = \sum_{\rho \sigma} \ket{\rho \sigma; \bs{M}(t)} \bra{\rho \sigma} + \sum_{\lambda \sigma} \ket{\lambda \sigma} \bra{\lambda \sigma}$, 
where $\ket{\rho \sigma; \bs{M}(t)}$ is an instantaneous eigenstate of $\bs{M}(t)\hat{\bs{\sigma}}$; formally, $\bs{M}(t)\hat{\bs{\sigma}} \ket{\rho \sigma; \bs{M}(t)} = M \sigma\, \ket{\rho \sigma; \bs{M}(t)}$ with $M= |\bs{M}|$. 
This transformation does not affect the left lead's Hamiltonian $\tilde H_l := U H_l U^\dagger = H_l$. But it diagonalizes the right lead's Hamiltonian $\tilde H_r := U H_r U^\dagger = \sum_{\rho \sigma} \ket{\rho \sigma} \xi_{\rho \sigma} \bra{\rho \sigma}$, where $\xi_{\rho \sigma}= \epsilon_\rho - M \sigma/2$. The magnetization's time-dependence is transferred to the tunneling Hamiltonian $\tilde H_t := U H_t U^\dagger = \sum_{\rho \lambda \sigma \sigma'} \ket{\rho \sigma} [R^\dagger(t)]_{\sigma \sigma'} t_0 \bra{\lambda \sigma'} + h.c.\ $. That is, the tunneling amplitudes become time-dependent and non-trivial in spin space,
\begin{equation}
t_0 \rightarrow R^\dagger(t)\, t_0\ . \label{eq: tunneling elements}
\end{equation}
The spin-space rotation $R(t)$ is defined by its elements $[R(t)]_{\sigma \sigma'} := \braket{\rho \sigma; \bs{M}(t)}{\rho \sigma'}$. 
Note, however, that this definition is not unique, as a rotation around the magnetization direction (spin quantization axis) has no physical effect. This gives rise to a gauge freedom which can be used to simplify the calculation \citep{PhysRevLett.114.176806}.

Due to its time dependence, the transformation not only rotates the Hamiltonian but it also generates a new term, $-i U \dot U^\dagger = - i \sum_{\rho \sigma \sigma'}\ket{\rho \sigma} [R^\dagger \dot R]_{\sigma \sigma'} \bra{\rho \sigma'}$, in the rotating-frame Hamiltonian. The spin-off-diagonal part of $-i R^\dagger \dot R$ induces transitions---also known as Landau-Zener-transitions---between spin-up and spin-down states. However, we assume a large magnetization length $M$, such that we can disregard these transitions \footnote{Via $\dot R$, the term $-i R^\dagger \dot R$ is related to the dynamics of the magnetization. So, it becomes smaller for slower magnetization dynamics. Explicitly, this can be seen from the spin-off-diagonal elements $-i [R^\dagger \dot R]_{\sigma \bar \sigma} = \frac{\dot \phi \sin \theta}{2} e^{-i \sigma  \dot \phi \cos \theta\, t}$, where $\bar \sigma$ denotes the spin opposite to $\sigma$. To disregard these spin-off-diagonal terms, they must be small compared to the spin-diagonal elements. In more physical terms, the magnetization must be slow compared to its length.}; this is also known as adiabatic approximation \cite{PhysRevLett.114.176806}. The remaining spin-diagonal part of $-i R^\dagger \dot R$ gives an additional time-evolution phase---also known as Berry-phase---which is different for spin-up and spin-down states. However, we eliminate the spin-diagonal part of $-i R^\dagger \dot R$ by fixing the gauge analog to Ref. \cite{PhysRevLett.114.176806}; that is, we explicitly choose
\begin{equation}
R(t) = \mat{\cos \frac{\theta}{2}\ e^{-i \omega_- t}}{-\sin \frac{\theta}{2}\ e^{-i \omega_+ t}}{ \sin \frac{\theta}{2}\ e^{i \omega_+ t}}{ \cos \frac{\theta}{2}\ e^{i \omega_- t}}\ ,\label{eq: rotation}
\end{equation}
with $\omega_\pm = \dot \phi (1 \pm \cos \theta)/2$, where $\omega_-$ is the rate at which the Berry-phase is acquired. Even though this choice eliminates the spin-diagonal part of $-i R^\dagger \dot R$, it does not eliminate the Berry-phase. Instead, the Berry-phase is transferred to the tunneling elements, Eq. \eqref{eq: tunneling elements}. To summarize: for the specific choice, Eq. \eqref{eq: rotation}, the newly generated term $-i R^\dagger \dot R$ can be disregarded in an adiabatic approximation.

Now, in the rotating frame, it is straightforward to apply Fermi's golden rule. Treating the tunneling Hamiltonian $\tilde H_t$ as perturbation, we obtain the golden-rule rate 
\begin{align}
\Gamma_{\lambda \sigma' \rightleftarrows \rho \sigma} = \frac{2 \pi}{\hbar} &|t_0|^2 \frac{1 + \sigma \sigma' \cos \theta}{2} 
 \nonumber \\  &\times \delta \left( \xi_{\rho \sigma} - \epsilon_\lambda + \sigma' \hbar \omega_{\sigma \sigma'} - eV \pm E_c \right)\ , \label{eq:FGR}
\end{align}
where $+E_c$ and $-E_c$ correspond to left-to-right tunneling $\lambda \sigma' \rightarrow \rho \sigma$ and right-to-left tunneling $\rho \sigma \rightarrow \lambda \sigma'$ respectively. 
The \textit{macrospin orientation} governs the spin-projection factor $(1+ \sigma \sigma' \cos \theta)/2$, which is $\cos^2 \frac{\theta}{2}$ for equal spins $\sigma = \sigma'$ and $\sin^2 \frac{\theta}{2}$ for opposite spins $\sigma \neq \sigma'$. 
The \textit{macrospin dynamics} enters through the frequency $\omega_{\sigma \sigma'} = \dot \phi (1 - \sigma \sigma' \cos \theta)/2$, which is $\omega_-$ for equal spins $\sigma = \sigma'$ and $\omega_+$ for opposite spins $\sigma \neq \sigma'$. 
In the rotating frame, the macrospin dynamics translates into the time-dependence of the perturbation (tunneling Hamiltonian); see Eqs. \eqref{eq: tunneling elements} and \eqref{eq: rotation}. Consequently, the macrospin dynamics induces the energy shift, $\sigma' \hbar \omega_{\sigma \sigma'}$, in the golden-rule rate. In other words, the precessing macrospin gives energy to---or takes energy from---the tunneling electrons; that is, it can assist electrons in tunneling \cite{ludwig2019current}.

Now, knowing the golden-rule rate, Eq. \eqref{eq:FGR}, we can determine the charge current.

\section{Charge current}The net charge current $I= I_{l \rightarrow r} - I_{r \rightarrow l}$ is the difference between the left-to-right current $I_{l \rightarrow r}$ and the right-to-left current $I_{r \rightarrow l}$.

\begin{figure}[t]
\begin{center}
\includegraphics[width=0.475\textwidth]{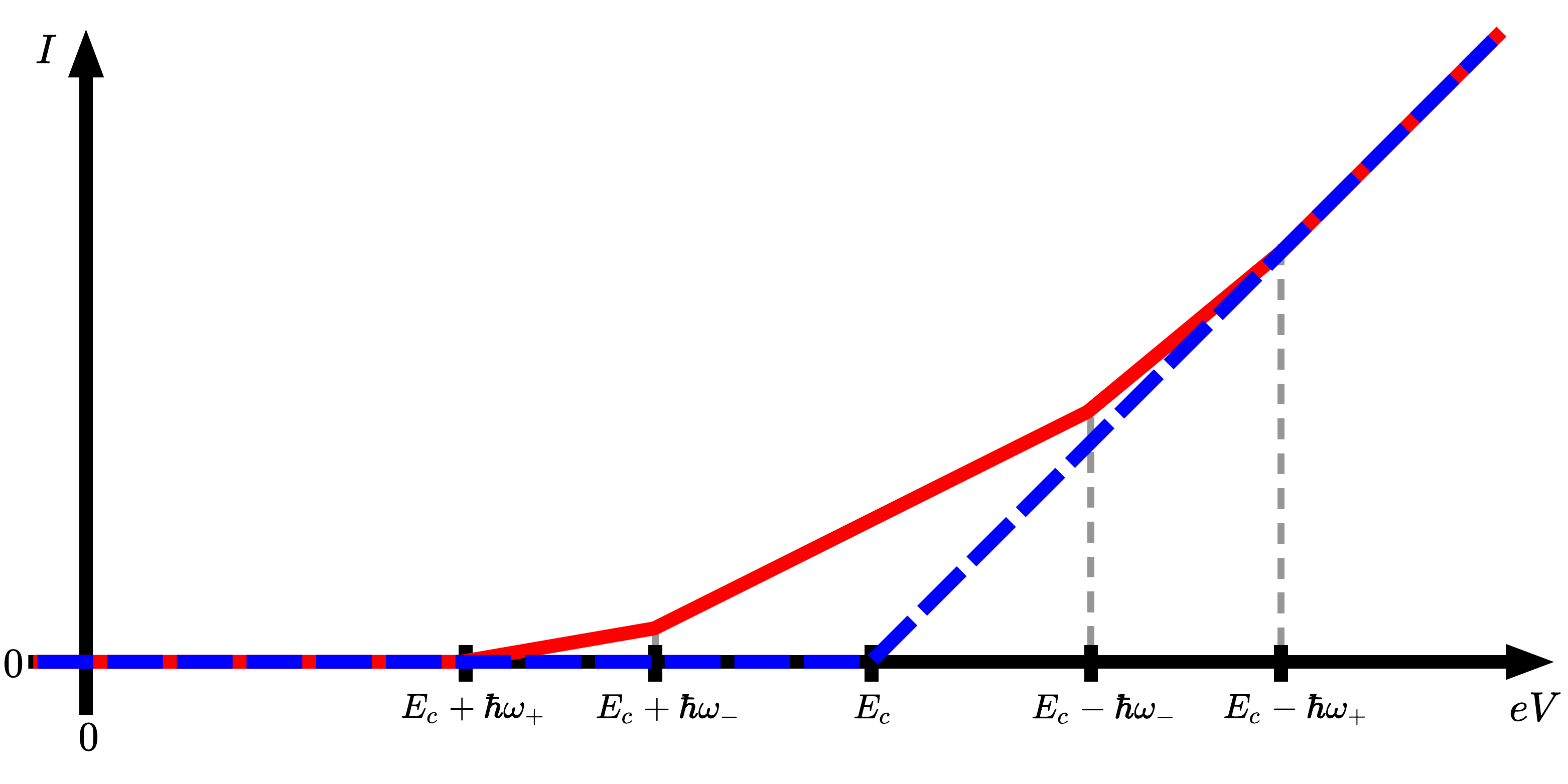}
\end{center}
\caption{This figure shows the softening of Coulomb blockade for slow precession ($\hbar |\dot \phi| < E_c$) with $0<\theta< \pi/2$ and $\dot \phi <0$. 
While charge transport is still blocked for low voltages; due to macrospin-assisted tunneling, a current flows already for $eV> E_c + \hbar \omega_+$. The details of the current flow depend on the macrospin dynamics. Thus, the macrospin dynamics can be measured (indirectly) by the charge current. The standard Coulomb blockade is included as reference (blue-dashed).} \label{fig: softening}
\end{figure}

At first, we focus on the left-to-right current $I_{l \rightarrow r}$; that is, we consider only electrons that are tunneling from the left lead to the right lead. Formally, it is given by $I_{l \rightarrow r} = e  \sum_{\rho \lambda \sigma \sigma'} \Gamma_{\lambda \sigma' \rightarrow \rho \sigma} f(\epsilon_\lambda) [1- f(\xi_{\rho \sigma})]$, where the golden-rule rate, Eq. \eqref{eq:FGR}, is summed over all states and---since electrons can only tunnel from filled states into empty states---it is weighted by the filling factor $f(\epsilon_\lambda)$ and the Pauli-blocking factor $[1- f(\xi_{\rho \sigma})]$. More explicitly,
\begin{align}
&I_{l \rightarrow r}\! =\! \frac{g_t}{2e} \sum_{\sigma \sigma'}  \frac{1\! +\! \sigma \sigma'\! \cos \theta}{2} \label{eq: left-to-right current} \\
 &\times \!\! \int\!\! d \epsilon_l\!\! \int\!\! d \epsilon_r\, f(\epsilon_l\! -\! eV) [1\! -\! f(\epsilon_r)]  \delta(\epsilon_r\! -\! \epsilon_l\! +\! E_c\! +\! \sigma' \hbar \omega_{\sigma \sigma'})\ , \nonumber
\end{align}
where we shifted the integrals $\epsilon_l \rightarrow \epsilon_l - e V$ and $\epsilon_r \rightarrow \epsilon_r + M\sigma /2$. Furthermore, we assumed the densities of states $\rho_l$ and $\rho_r$ to be constant on all scales smaller than $M$ and to be independent of the spin \footnote{Formally, the spin independence means $\rho_r(\epsilon - M/2) = \rho_r(\epsilon + M/2) = \rho_r$}. The tunneling conductance $g_t$ is defined by $g_t =  8 \pi^2 |t_0|^2 \rho_l \rho_r\, e^2/h$. From Eq. \eqref{eq: left-to-right current}, it becomes clear that $eV$ is just the electrical part of the electrochemical potential:  $f(\epsilon_l - eV)  = 1/[\exp[(\epsilon_l - \mu_l)/T] +1] $, where $\mu_l = \mu + eV$ is the electrochemical potential of the left lead.

For infinite capacitance ($E_c=0$) and without magnetization dynamics ($\omega_{\sigma \sigma'}=0$), the $\delta$-function in Eq. \eqref{eq: left-to-right current}  ensures the conservation of energy for the tunneling electrons. For finite capacitance, however, a tunneling electron loses the charging energy $E_c$ to the electrostatic environment (capacitor) \cite{ingold1992charge, nazarov2009quantum}; see Fig. \ref{fig: system}. The energy shift $\sigma' \hbar \omega_{\sigma \sigma'}$ accounts for the effect of the macrospin dynamics onto the tunneling electron; namely, it describes the energy gain or loss due to the macrospin precession. Performing the integrals in Eq. \eqref{eq: left-to-right current}, we obtain 
\begin{align}
I_{l \rightarrow r}= \frac{g_t}{2e} \Big[ &\cos^2 \frac{\theta}{2}\ \Pi  (eV\! -\! E_c\! -\!  \hbar \omega_-)   \nonumber \\
+  &\sin^2 \frac{\theta}{2}\ \Pi (eV\! -\! E_c\! +\!  \hbar \omega_+)  \nonumber \\
+ &\sin^2 \frac{\theta}{2}\ \Pi  ( eV\! -\! E_c\! -\!  \hbar \omega_+ )  \nonumber \\ + &\cos^2 \frac{\theta}{2}\ \Pi  ( eV\! -\! E_c\! +\!  \hbar \omega_- )\Big]\ , \label{eq: exlicit left-to-right current}
\end{align}
where $\Pi(x)$ is the ramp function; that is, $\Pi(x)=0$ for $x\leq 0$ and $\Pi(x) = x$ for $x>0$. The four terms in $I_{l \rightarrow r}$ arise from the different combinations of spins in left-to-right tunneling.

The right-to-left current $I_{r \rightarrow l}$ can be found analogously to the left-to-right current $I_{l \rightarrow r}$; only the roles of the leads are exchanged \footnote{Formally, we have $I_{r \rightarrow l} = e \sum_{\rho \lambda \sigma \sigma'} \Gamma_{\rho \sigma \rightarrow \lambda \sigma'} f(\xi_{\rho \sigma}) [ 1 - f(\epsilon_\lambda)]$. Explicitly, we obtain $I_{r \rightarrow l} = \frac{g_t}{2e} \big[\cos^2 \frac{\theta}{2}\ \Pi (-eV\! -\! E_c\! +\!  \hbar \omega_-)  + \sin^2 \frac{\theta}{2}\ \Pi  (-eV\! -\! E_c\! -\!  \hbar \omega_+)  + \sin^2 \frac{\theta}{2}\  \Pi ( -eV\! -\! E_c\! +\!  \hbar \omega_+ ) +\cos^2 \frac{\theta}{2}\  \Pi(-eV\! -\! E_c\! -\!  \hbar \omega_- ) \big]$.}.
Combining both, we find that the charge current, $I = I_{l \rightarrow r} - I_{r \rightarrow l}$, is antisymmetric in the voltage; that is, $I(-V) = - I(V)$. This is a consequence of assuming the densities of states to be spin-independent.

To gain a better understanding of the charge current, let's consider a situation with static macrospin ($\dot \phi = 0$) at first. In the limit of infinite capacitance ($E_c = 0$), the tunnel-junction behaves as a resistor; that is, the current-voltage relation is described by Ohm's law $I = g_t V$. For finite capacitance ($E_c >0$), in contrast, the tunneling electrons loose the energy $E_c$ to the electrostatic environment. This loss effectively reduces the voltage by $E_c/e$. Consequently, we obtain $I = g_t [(V - E_c /e)\, \Theta (V - E_c /e) - (-V - E_c /e)\, \Theta (-V - E_c /e)]$, which is the standard Coulomb blockade result \cite{ingold1992charge, nazarov2009quantum}: if the voltage is too low ($|eV|<E_c$), the charge transport is blocked ($I=0$). However, when the macrospin precesses ($\dot \phi \neq 0$), it can assist electrons in tunneling; thereby, it softens the Coulomb blockade or---if the precession is strong enough---it can even break the Coulomb blockade.

\section{Breaking of Coulomb blockade}
When the macrospin precesses slowly ($\hbar |\dot \phi| < E_c$), the Coulomb blockade is softened: electrons can tunnel through the junction, even if the applied voltage is smaller than---but close enough to---the charging energy; see Fig. \ref{fig: softening}. The missing energy is provided by the precessing macrospin. So, the macrospin dynamics governs the softening of Coulomb blockade. In turn, a measurement of the charge current can reveal the macrospin dynamics. For example in the voltage regime $E_c + \hbar \omega_+ < eV < E_c + \hbar \omega_-$, compare Fig. \ref{fig: softening}, the current is given by $I = g_t \sin^2(\theta/2) [ eV - E_c  - \hbar \dot \phi \cos^2(\theta/2) ]/2e$. Thus, a measurement of the differential conductance $\frac{dI}{dV} =  \sin^2(\theta/2) g_t/2$ reveals the polar angle $\theta$. Then, knowing $\theta$, the precession frequency $\dot \phi$ can be inferred from the current $I$ itself. A shortcoming of this method is that one has to know in which regime the voltage is. A simpler way would be to measure the current-voltage characteristics, Fig. \ref{fig: softening}, and determine the magnetization dynamics from the position of the kinks at $E_c \pm \hbar \omega_+$ and $E_c \pm \hbar \omega_-$---or analogously from the position of jumps in the differential conductance.

\begin{figure}[t]
\begin{center}
\includegraphics[width=0.475\textwidth]{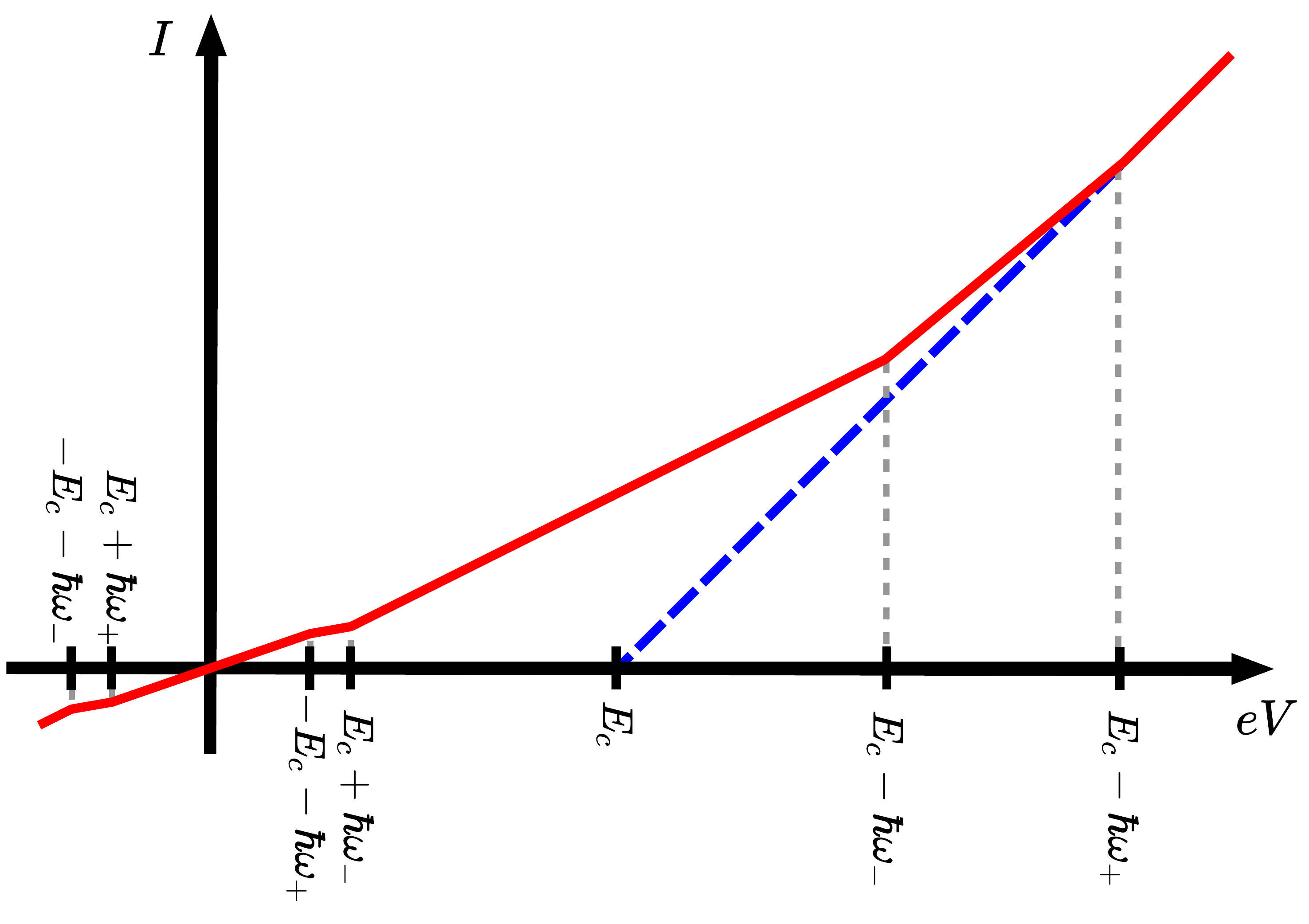}
\end{center}
\caption{This figure shows the breaking of Coulomb blockade for strong macrospin precession ($\hbar |\omega_+| > E_c$) with $0<\theta< \pi/2$ and $\dot \phi <0$. The standard Coulomb blockade is included as reference (blue-dashed). Because of the macrospin-assisted tunneling, the Coulomb blockade disappears; that is, a charge current flows at arbitrary low (but nonzero) voltages. Yet, the details depend on the macrospin dynamics. In turn, a measurement of charge current can reveal the macrospin dynamics.} \label{fig: breaking}
\end{figure}

While only softened for slow precession, the Coulomb blockade is completely broken for strong macrospin precession ($\hbar |\omega_+|>E_c$ and/or $\hbar |\omega_-|>E_c$). In this case, the precessing macrospin gives enough energy to the tunneling electrons, such that tunneling is possible even if there is no other source of energy. In turn, even at low voltages, we find a linear relation between current and voltage; see Fig. \ref{fig: breaking}. So, the macrospin dynamics governs the breaking of Coulomb blockade. And again, a measurement of the charge current can reveal the macrospin dynamics. However, in contrast to the softening of Coulomb blockade, the (differential) conductance can reveal the polar angle $\theta$ even at zero voltage. For example in the low voltage regime ($E_c +\hbar \omega_+ < eV < - E_c - \hbar \omega_+$), as shown in Fig. \ref{fig: breaking}, the current is given by $I = g_t V \sin^2 (\theta/2)$, which reveals the polar angle $\theta$ but not the precession frequency $\dot \phi$ \footnote{Note, however, that this only works if $|\hbar \omega_+| > E_c$ and $|\hbar \omega_-| < E_c$. If $|\hbar \omega_+| > E_c$ and $|\hbar \omega_-| > E_c$, then we would simply find Ohm's law $I = g_t V$ at low voltages.}. To determine the precession frequency, one has to go to higher voltages again.

\section{Discussion}
We have found that macrospin-assisted tunneling can break the Coulomb blockade. More explicitly, we considered a tunnel junction between a normal metal and a metallic ferromagnet where the macrospin (magnetization) is in a steady state precession. The precessing macrospin creates a time-dependent field for electrons, which can assist them in tunneling \cite{ludwig2019current}. As we have shown, this macrospin-assisted tunneling shrinks the regime of Coulomb blockade; see Fig. \ref{fig: softening}. When the macrospin precession is strong enough, the regime of Coulomb blockade vanishes completely; see Fig. \ref{fig: breaking}. In other words, the macrospin-assisted tunneling can soften or even break the Coulomb blockade. The details of the softening or breaking of Coulomb blockade depend on the macrospin dynamics. Thus, a measurement of the charge current can reveal the macrospin dynamics.

To get a better understanding of the scales involved, let's consider a specific system. For example in Ref. \cite{cascales2015detection}, they report on a magnetic tunnel-junction with elliptical shape (minor axis $40\, \mathrm{nm}$; major axis $80\, \mathrm{nm}$) and a MgO tunnel-barrier with thickness $0.9\, \mathrm{nm}$. This leads to a capacitance of $C \approx 0.25\, \mathrm{fF}$ for the tunnel-junction \footnote{To estimate the capacitance, we used the formula for parallel plates $C = \epsilon_0 \epsilon_r A / d$; with the dielectric constant $\epsilon_0 \approx 9 \cdot 10^{-12} \frac{\mathrm{F}}{\mathrm{m}}$, the relative permittivity for $\mathrm{MgO}$ which is $\epsilon_r \approx 10$, the ellipsis area $A = \frac{\pi}{4} 40 \cdot 80 \mathrm{nm}^2$, and the barrier thickness $d = 0.9\, \mathrm{nm}$}. In turn, we find a charging energy of $E_c \approx 0.32\, \mathrm{meV}$ which corresponds to a temperature of $T_c = E_c/k_B \approx 3.7\, \mathrm{K}$ and a frequency of $f_c = E_c/h \approx 78\, \mathrm{GHz}$. To enter the regime of Coulomb blockade, the temperature must be well below $T_c$. Then, the precessing macrospin could break the Coulomb blockade, if  it precesses at frequencies above $f_c$. While the precession frequency reported in Ref. \cite{cascales2015detection} is only of the order of $10\, \mathrm{GHz}$, it is still close enough to the critical frequency $f_c$, such that one can expect a clear softening of the Coulomb blockade; analog to Fig. \ref{fig: softening}. For a tunnel-junction of larger dimensions and with a thinner barrier, the critical frequency $f_c$ can fall below $10\, \mathrm{GHz}$ such that one might also observe the breaking of Coulomb blockade; analogous to Fig. \ref{fig: breaking}.

Also beyond the specific setup considered here, the breaking of Coulomb blockade by macrospin-assisted tunneling might be interesting; in particular, for scanning tunneling microscope (STM) setups \cite{PhysRevB.90.174431}. In STM setups, the capacitance is harder to estimate; see Ref. \cite{de2017fast} for example. However, in Ref. \cite{PhysRevLett.124.156803}, where they also use the Coulomb blockade to investigate a system in a scanning tunneling spectroscopy (STS) setup, they find a junction capacitance of $C = 21.7\, \mathrm{fF}$. This capacitance corresponds to a charging energy of $E_c \approx 3.7\, \mathrm{\mu eV}$, a temperature of $T_c = E_c/k_B \approx 42\, \mathrm{mK}$, and a frequency of $f_c = E_c/h \approx 0.9\, \mathrm{GHz}$. So, in this case, a macrospin precession frequency of roughly $10\, \mathrm{GHz}$ would be well above $f_c$, such that the macrospin assisted tunneling can easily break the Coulomb blockade. This effect might be particularly interesting for resonant-state-STM setups \cite{PhysRevLett.123.087202, schlenhoff2020real}---where the charging energy $E_c$ can be tuned, because of a large variability in the distance between STM-tip and probe material.

While we focused on a passive use (indirect measurement of magnetization dynamics), we can also think of more active uses of the interplay between Coulomb blockade and macrospin-assisted tunneling. It could be used to control a spin current electrically \footnote{For example, for the softening of Coulomb blockade, Fig. \ref{fig: softening}, we have a purely spin-polarized current in the regime $E_c + \hbar \omega_+ < eV < E_c + \hbar \omega_-$. In this regime, electrons tunnel only from spin-up states to spin-up states. And the amount of tunneling electrons (the current) can be controlled by the applied voltage.}. Or, when the magnet's density of states is spin-dependent, it can be used to pump a charge current \footnote{The charge pumping is similar to Ref. \cite{PhysRevB.78.020401} but, interestingly, it works for a tunnel-junction between a magnet and a normal metal; that is, without a second magnet.}. Because it can be used to control spin and charge currents, it might also open up new ways to control the magnetization dynamics. From a more general perspective, the interplay between Coulomb blockade and macrospin-assisted tunneling provides a new platform for the interplay between electronics and spintronics. From this perspective, magnon-assisted tunneling (as considered in \cite{zhang1997quenching, bender2019quantum, han2001analyses, lu2003magnon, balashov2008inelastic}) is a natural candidate for the generalization of our results.

\section*{Acknowledgements}
We thank I. S. Burmistrov, P. Simon, M. Trif, and W. Wulfhekel for a fruitful discussion. This work is part of the research programme Fluid Spintronics with project number 182.069, which is (partly) financed by the Dutch Research Council (NWO).

\bibliography{reference.bib}

\end{document}